\documentclass[11pt, conference]{IEEEtran}
\usepackage{subfigure}
\usepackage{caption}
\usepackage{array}
\usepackage{lipsum}
\usepackage[]{algorithm2e}
\usepackage{tabularx}
\usepackage{hyperref}

\usepackage{graphicx}
\usepackage{booktabs}
\usepackage{float}
\usepackage{multirow}
\usepackage{multicol}
\usepackage{array}
\usepackage{tabularx}
\usepackage{bm}
\usepackage{soul}
\usepackage{tikz}
\usepackage{amssymb}
\usepackage{amsmath}
\usepackage{xfrac}
\usepackage[bottom]{footmisc}
\usepackage{fixltx2e}
\begin{document}
\title{FPGA Acceleration of Sequence Alignment: A Survey}
\author {Sahand Salamat, Tajana Rosing \\
Department of Computer Science and Engineering, UC San Diego, La Jolla, CA 92093, USA\\
  \{sasalama, tajana\}@ucsd.edu}
\maketitle

\begin{abstract}
Genomics is changing our understanding of humans, evolution, diseases, and medicines to name but a few.  As sequencing technology is developed collecting DNA sequences takes less time thereby generating more genetic data every day. Today the rate of generating genetic data is outpacing the rate of computation power growth. Current sequencing machines can sequence 50 humans genome per day; however, aligning the read sequences against a reference genome and assembling the genome will take 1300 CPU hours \cite{li2013aligning}. The main step in constructing the genome is aligning the reads against a reference genome. Numerous accelerators have been proposed to accelerate the DNA alignment process. Providing massive parallelism, FPGA-based accelerators have shown great performance in accelerating DNA alignment algorithms. Additionally, FPGA-based accelerators provide better energy efficiency than general-purpose processors. In this survey, we introduce three main DNA alignment algorithms and FPGA-based implementation of these algorithms to accelerate the DNA alignment. We also, compare these three alignment categories and show how accelerators are developing during the time.
 
\end{abstract}

\section{Introduction} \label{sec:intro}
The first human genome assembled in 2001 \cite{international2001initial} since then the size of the genomics data has been doubled every 7 months \cite{turakhia2018darwin} which is significantly higher comparing to Moore's Law which expect to double the computation power every 2 years. The improvement in genome sequencing tools led to this massive data generation rate. Genome sequencing tools read the sequence of nucleotide bases in DNA molecules. The genomics data is valuable to find the DNA mutations causing cancer \cite{pleasance2010comprehensive}, Alzheimer \cite{lacour2017genome}, genetic disorders \cite{cho2017prevalence}, and autism \cite{krumm2015excess} to name but a few. This data can also be used to understand the humans and their differences \cite{spichenok2011prediction,sequencing2005initial, mclean2011human} better to create personalized medicine for each person and each disease. \cite{hamburg2010path}. 

Sequencing is the process of identifying the nucleotides in the DNA sequence. DNA sequencing machines output fragments of the DNA referred to as \textit{reads} \cite{al2016survey}. The first generation of DNA sequencing is deployed in the 1990s which resulted in assembling the human genome for the first time with a cost of 3 million dollars. These machines mostly relied on the \textit{Sanger Sequencing} method introduced in the late 70s\cite{sanger1977dna}. The technique reads fragments of the DNA sequence with a few thousand base pairs long. These reads are useful for sequencing unknown genomes where no reference genome is available. This process requires high cost and long time to sequence DNA thereby limiting the usage of these sequencers. 

In the next generation sequencing (NGS) technology the long and slow reads were replaced with short and massively parallel reads. This technology outputs reads with 100 base pairs to 1000 base pairs \cite{metzker2010sequencing}. One Illumina’s HiSeq machine can read up to 45 human genomes per day \cite{kumar2015metagenomics}, or a portable sequencer device can sequence a human genome in several days \cite{eisenstein2012oxford}. This sequencing technology dropped the sequencing cost to a thousand dollars per genome \cite{fujiki2018genax}.

Third generation sequencing machines generate much longer reads in the order of $\sim$10000 bases \cite{schadt2010window}. Longer reads make assembling genomes significantly faster and more accurate \cite{gordon2016long}. For personalized medicine, and understanding diseases, longer reads can identify long insertions and deletions. Moreover, a DNA sequence would be divided into fewer fragments which makes assembling easier \cite{schadt2010window}. However, the main disadvantage of the third generation sequencing machines is their high error rate, $\sim15\%$ to $\sim40\%$ in the sequencing process\cite{goodwin2015oxford}.

Sequencing machines produce multiple reads in each run. The next step to process the generated data is finding the locations where the reads are similar to the reference genome, called read alignment, or assembling the reads next to each other to generate the genome. Several algorithms have been proposed to align the read sequences against the reference genome. However, as the amount of data generated every day increases, general-purpose processors have become incapable of processing all the generated data. Therefore, application specific accelerators \cite{turakhia2018darwin, salamat2017systematic} and FPGA-based accelerators have been introduced to increase the performance and efficiency of processing genomics data in general and DNA alignment in particular due to their reconfigurability along with their highly parallel architecture. In this paper, we first introduce the most common DNA alignment algorithms. Then in the following sections, we introduce the FPGA-based accelerators for each category of the alignment algorithms. In the end, we compare the advantages and disadvantages of each algorithm and conclude the paper. 

\section{Background and Algorithms}
\subsection{FPGAs}
\begin{figure}[t]
  \centering
  {\includegraphics[width=1\linewidth]{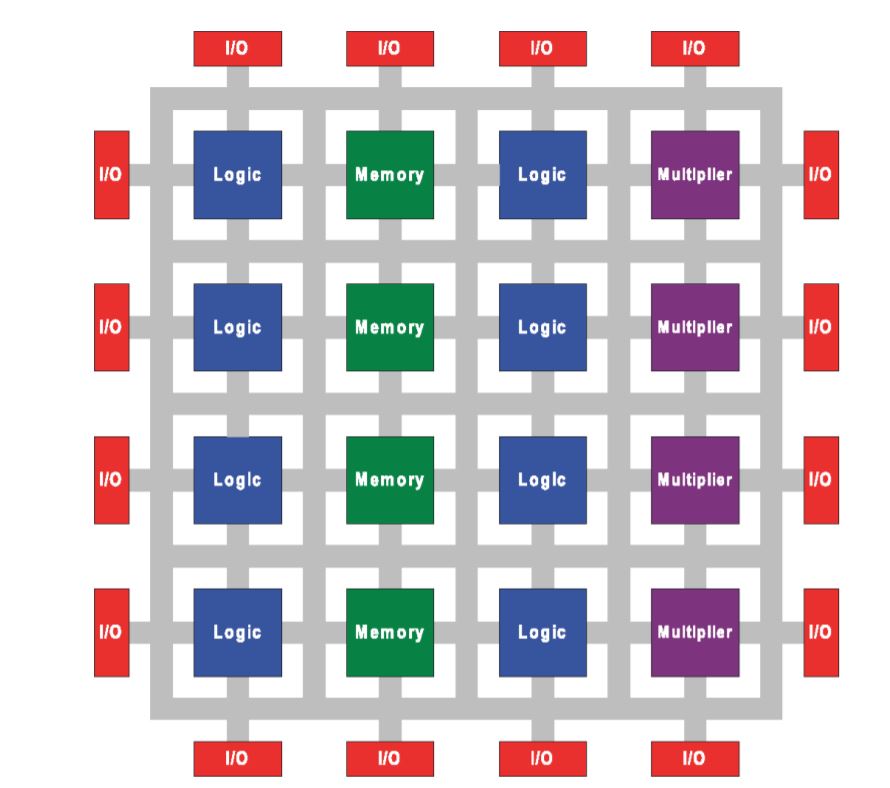}}
  \caption{Architecture and available resources in FPGAs\cite{kuon2008fpga}}
  \label{fig:fpga_arch}
\end{figure}

Field-Programmable Gate Arrays (FPGAs) are programmable devices that can be configured to become almost any kind of digital design. Figure \ref{fig:fpga_arch} shows the architecture of an FPGA which consists of an array of programmable logic blocks. There are different types of programmable blocks including Look-Up Tables (LUTs), which are the main resources to implement any logic on FPGAs, memory blocks, and Digital Signal Processing blocks (DSPs), surrounded by a programmable routing fabric that brings programmability to the interconnects. Programmable input and output blocks are also provided for off-chip communication. The hardware described in Hardware description Languages (HDLs) are synthesized and mapped to the FPGA resources, and then programmed on the FPGA \cite{salamat2019workload}. During the FPGA programming the content of the LUTs and routing modules are set. Reconfigurablity, and programability of FPGAs have made them a great platform to prototype and accelerate various type of applications from machine learning \cite{salamat2019f5, imani2019sparsehd, salamat2018rnsnet, imani2019quanthd} to bionformatics. DNA alignment algorithms in specific, require irregular parallelism and they can benefit greatly from the FPGA parallelization. During the past decade, many works have tried to accelerate these algorithms on FPGAs. In the next sections some of the most promising works on accelerating DNA alignment is introduced.

\subsection{Pairwise Sequence Alignment}
 Several algorithms have been proposed for sequence alignment, one way to align two sequences is by using \textit{Pairwise Sequence Alignment} (PSA). In this method, elements of the query sequence are compared and aligned against the reference sequence. PSA is usually implemented using dynamic programming algorithms, such as the Needleman-Wunsch (NW) \cite{needleman1970general}, and Smith-Waterman (SW) \cite{smith1981identification}. These two algorithms compare elements of two sequences and assign a positive score if the elements are matched and a negative penalty for mismatches, insertion, and deletion. The penalties for mismatches, deletion, and insertion may vary due to biological reasons. These algorithms then use dynamic programming to find the highest score which represents the alignment with the highest similarity.
 
\begin{figure}[t]
  \centering
  {\includegraphics[width=1\linewidth]{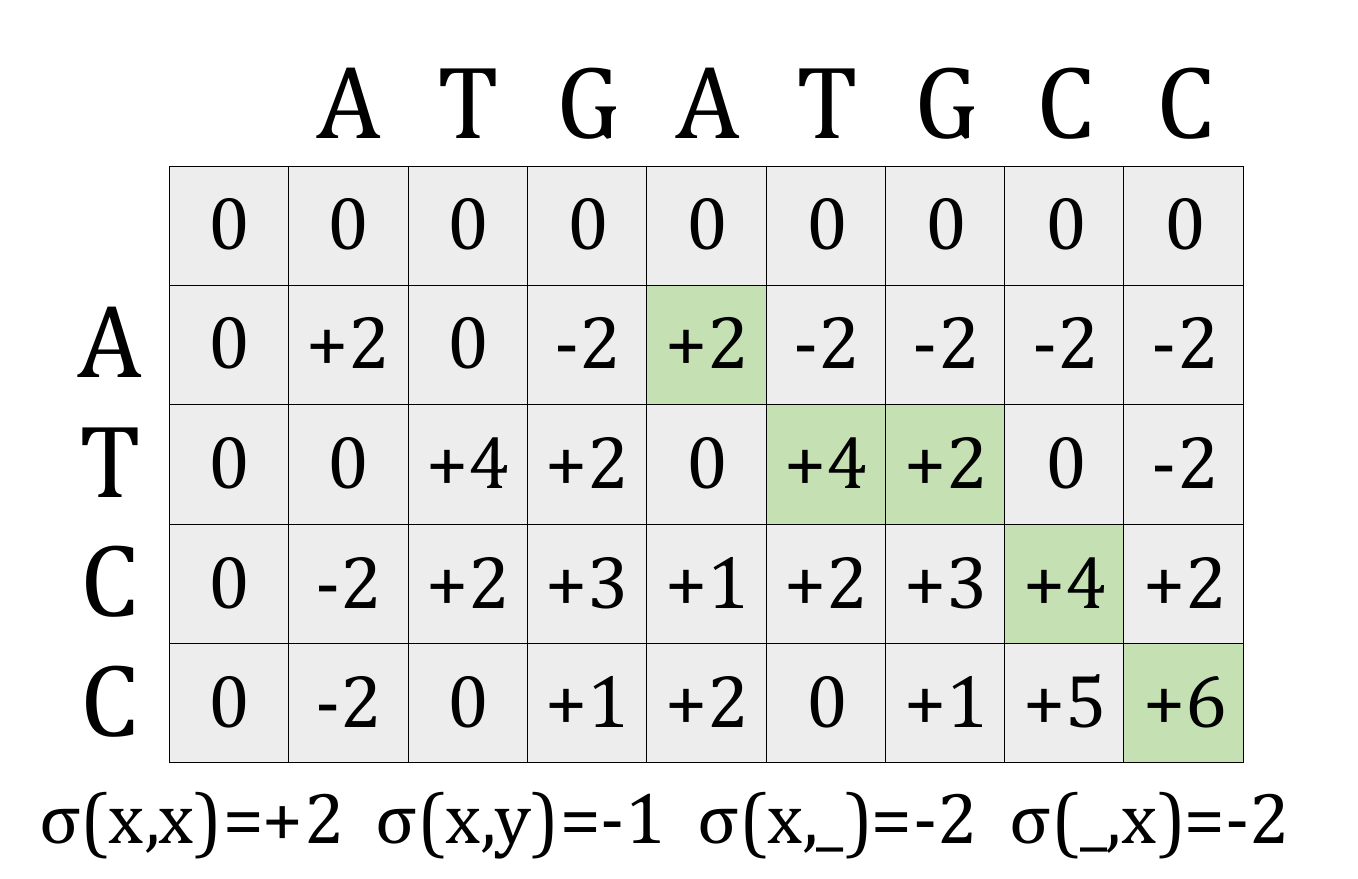}}
  \caption{H matrix for SW and NW algorithms calculated using dynamic programming}
  \label{fig:SW}
\end{figure}

 The main difference between NW and SW algorithm is that the NW algorithm globally aligns two sequences while SW finds the local alignment. These algorithms are used to find the optimal alignment between the reference sequence (S) and the query sequence (Q). The reference sequence has a length of $|S|=n$ and the query sequence has a length of $|Q|=m$. A matrix with size $m+1 \times n+1$ called the alignment matrix ($H$) is used to describe how $S$ and $Q$ sequences align against each other. The $S$ sequence is placed at the first row, while the $Q$ sequence is placed at the first column. The value of each element of the matrix, which shows the score of alignment, is calculated using dynamic programming. Equation \eqref{eq:SW-B} shows the boundary conditions of the matrix H and equation \eqref{eq:SW-DP} shows the recursive equation used for calculating $H (i, j)$ where $1 \leqslant  i \leqslant m, 1\leqslant j \leqslant n $.
\small{
\begin{subequations}\label{eq:rd}
\begin{align}\label{eq:SW-B}
\begin{cases}
H(i, 0) = 0 \quad  for \ 0 \leqslant  i \leqslant m \\
H(0, j) = 0 \quad  for \ 0 \leqslant  i \leqslant n \\
\end{cases}
\end{align}
\begin{align}\label{eq:SW-DP}
H(i,j)= max
\begin{cases}
0 \hspace{28pt}(Only\ SW)\\ 
H(i-1, j-1)+ \sigma(S[i], Q[j])\   (Mis)Match \\
H(i-1, j)+ \sigma(S[i],\_) \hspace{35pt} Deletion \\
H(i, j-1)+ \sigma(\_, Q[j]) \hspace{33pt} Insertion \\
\end{cases}
\end{align}
\end{subequations}
}
The function $\sigma(x, y)$ determines the positive score of matches, and penalties for mismatches, insertions, and deletions. The scores/penalties can be adjusted according to the biological inspirations (e.g. deletions and insertions are less likely than mismatches) \cite{al2016survey}. Figure \ref{fig:SW} shows the calculated matrix $H$ for a reference sequence \textit{S="ATGATGCC"} and a query sequence \textit{Q="ATCC"}. Scores and penalties are also defined as $\sigma(x, y)$ in the figure. In matrix $H$ the highest score indicates the most similar pattern in the reference sequence to the query sequence. By backtracking from the highest score to element with 0 value, we can find the optimal alignment, In this example, the query sequence is aligned with $AT\_CC$ from the reference sequence.

\subsection{Hash Tables}
Although PSA algorithms can find the optimum alignment, they are computationally intensive. Therefore, other algorithms, targeted heuristic approaches to reduce the computation complexity while delivering a desirable alignment. In this subsection we introduce a heuristic method based on hash tables it is worth mentioning that other hash-table based algorithms will be explained briefly in section \ref{sec:fpga_hash}. The work \cite{ning2001ssaha}, SSAHA, searches for exact or partial occurrences of a query sequence in reference sequences $\{S_1, S_2, \dots\}$ by using hash tables. SSAHA breaks each sequence into consecutive k-tuples represented in a hash table with the position of its occurrence in the sequence. A k-tuple is a $k$ bases long contiguous sequence. A sequence with $n$ bases, $|S_i|=n$, contains $(n-k+1)$ overlapping k-tuples. Each k-tuple in a sequence is represented as it's offset with respect to the first base of the sequence. Therefore $W_j(S_i)$ represents the k-tuple with $j$ offset from the first base of the $i^{th}$ reference sequence $S_i$. 

The first step of the algorithm is constructing the hash table from the reference sequences. The hash table consists of $4^k$ rows, one row for each the k-tuples. In each row, the position of the occurrence of the k-tuple, in the reference sequence $i$ with $j$ offset, is stored as $(i,j)$. To construct the hash table, only non-overlapping k-tuples $\{W_0(S_i), W_k(S_i), W_{2k}(S_i), \dots\}$ are considered. For each of the $4^k$ possible k-tuples the position of its occurrences are stored in the hash table. We solve the alignment problem of the previous example using hash tables for k=2. In addition to the reference sequence of the previous example $(S_1)$, for the sake of clarity, we assume there is one more reference sequence in the database $(S_2)$. The reference sequences in the database are:\\
$S_1 \ = \ AATCCAGACT$ \\
$S_2 \ = \ TCGGTATCCGACTACGTC$\\
The rows of table \ref{tab:H1} show the possible $4^2=16$ 2-tuples and in each row, the position of the occurrence of each 2-tuple is listed. For searching the matches for the query sequence in the database, first, all the overlapping k-tuples are extracted. At the base $t$ all the positions of k-tuples in the database which are the same as the k-tuple starting with the base $t^{th}$ in the query sequence are listed $\{(i_1,j_1), \ (i_2,j_2), \dots, \ (i_r,j_r)\}$. Then from the list of positions the list of \textit{hits} is founded. The list shows how to align the query sequence with respect to the reference sequence.\\
$$\{ H_1=(i_1,j_1-t,j1), \ H_2=(i_2,j_2-t,j2), \dots,\ H_r=(i_r, j_r-t, j_r)\}$$\\
Then all the elements of the list of hits are added to the master list $M$ and sorted, first by the \textit{index} ($i$), then by the \textit{shift} ($j-t$), and \textit{offset} ($j$). The final step is finding \textit{runs} of hits in the M list, which are elements in the $M$ for which \textit{index} and \textit{shift} are identical. Insertion and deletion are taken into account by accepting a small difference in the \textit{shift} between consecutive hits in a run. 

Table \ref{tab:H2} shows the search for the query sequence of $Q\ =\ ATCCGAC$ in the database. The reference sequence has 7 bases; consequently, it has 6 overlapping 2-tuples. The last column of table \ref{tab:H2} shows the list $M$. For $S_1$, two hits have slightly different \textit{shifts}. Therefore, the blue elements show the approximate matching, which shows an insertion/deletion in the sequences. However, in the $s_2$the green elements show the run of three hits (with the same value for \textit{shift}), starting at the $7^{th}$ element of the $S_1$ and ending at the base $13^{th}$.
{\renewcommand{\arraystretch}{1.2} 
\begin{table}[]
\centering
\caption{Generated hash table for S1 and S2 sequences}
\label{tab:H1}
\begin{tabular}{lll}
\hline
W  & E(W) & Positions                         \\ \hline
AA & 0    & (1,1)                             \\
AC & 1    & (1,8) (2,11) (2,14)               \\
AG & 2    &  (1,6)                            \\
AT & 3    & (1,2) (2,6)                       \\
CA & 4    & (1,5)                             \\
CC & 5    & (1,4) (2,8)                       \\
CG & 6    & (2,2) (2,9)    (2,15)             \\
CT & 7    & (1,9) (2,12)                      \\
GA & 8    & (1,7) (2,10)                      \\
GC & 9    &                                   \\
GG & 10   & (2,3)                             \\
GT & 11   & (2,4) (2,16)                      \\
TA & 12   & (2,5)    (2,13)                   \\
TC & 13   & (1,3)    (2,1)    (2,7)    (2,17) \\
TG & 14   &                                   \\ 
TT & 15   &                                   \\ \hline
\end{tabular}
\end{table}
}

{\renewcommand{\arraystretch}{1.4} 
\begin{table}[]
\centering
\caption{List of matches for the query sequence}
\label{tab:H2}
\begin{tabular}{llll|l}
t & Wt(Q) & Positions & H & M \\ \hline
0 & AT & \begin{tabular}[c]{@{}l@{}}(1,2)\\ (2,6)\end{tabular} & \begin{tabular}[c]{@{}l@{}}(1,2,2)\\ (2,6,6)\end{tabular} & \multirow{6}{*}{  \begin{tabular}[c]{@{}l@{}}\textcolor{blue}{(1,2,2)}\\ \textcolor{blue}{(1,2,3)}\\ \textcolor{blue}{(1,2,4)}\\ (1,3,7)\\ (1,3,8)\\ (2,-1,2)\\ (2,0,1)\\ \textcolor{green}{(2,6,6)}\\ \textcolor{green}{(2,6,7)}\\ \textcolor{green}{(2,6,8)}\\ \textcolor{green}{(2,6,9)}\\ \textcolor{green}{(2,6,10)}\\ \textcolor{green}{(2,6,11)}\\ (2,9,14)\\ (2,12,15)\\ (2,16,17)\end{tabular}} \\ \cline{1-4}
1 & TC & \begin{tabular}[c]{@{}l@{}}(1,3)\\ (2,1)\\ (2,7)\\ (2,17)\end{tabular} & \begin{tabular}[c]{@{}l@{}}(1,2,3)\\ (2,0,1)\\ (2,6,7)\\ (2,16,17)\end{tabular} &  \\ \cline{1-4}
2 & CC & \begin{tabular}[c]{@{}l@{}}(1,4)\\ (2,8)\end{tabular} & \begin{tabular}[c]{@{}l@{}}(1,2,4)\\ (2,6,8)\end{tabular} &  \\ \cline{1-4}
3 & CG & \begin{tabular}[c]{@{}l@{}}(2,2)\\ (2,9)\\ (2,15)\end{tabular} & \begin{tabular}[c]{@{}l@{}}(2,-1,2)\\ (2,6,9)\\ (2,12,15)\end{tabular} &  \\ \cline{1-4}
4 & GA & \begin{tabular}[c]{@{}l@{}}(1,7)\\ (2,10)\end{tabular} & \begin{tabular}[c]{@{}l@{}}(1,3,7)\\ (2,6,10)\end{tabular} &  \\ \cline{1-4}
5 & AC & \begin{tabular}[c]{@{}l@{}}(1,8)\\ (2,11)\\ (2,14)\end{tabular} & \begin{tabular}[c]{@{}l@{}}(1,3,8)\\ (2,6,11)\\ (2,9,14) \end{tabular} &  \\
\end{tabular}%
\end{table}
}

\subsection{Burrows-Wheeler Transform with FM-Index}
Burrows-Wheeler Transform (BWT) is used in text compression due to its ability to transform regular strings into strings with long runs of the same character. Applying BWT, long sequences of the same character can be compressed using the run-length encoding. This lossless compression algorithm represents a string with characters and the number of its consecutive repetition. BWT is combined with a search technique based on FM-index \cite{ferragina2000opportunistic} to first compress the reference sequence, and then align a sequence against a reference sequence. 

BWT, in the first step, adds a special character \textit{\$} at the end of the string and then creates every possible rotation of the string in a table. For a sequence with length $l$, $l+1$ rows will be generated where each row is the sequence of the previous row with a rotational shift. Since all the possible rotations of the sequence are stored in the table, each column of the table has all the characters of the original sequence. In the second step, all the possible rotations (rows) are sorted lexicographically. BWT stores the last column of the table and proves by storing the last column the table can be reconstructed. The first column of the table can be reconstructed by sorting the last column lexicographically. It has been proven that by having the last column and first column of the BWT the whole table can be reconstructed. BWT with FM-index data structure can locate all the occurrences of a pattern in the reference sequence \cite{ng2017reconfigurable}. Therefore it has been widely used in many of the existing software aligners including Bowtie \cite{langmead2009ultrafast}, SOAP2 \cite{li2009soap2}, and BWA \cite{li2009fast}. 

\begin{figure}[t]
  \centering
  {\includegraphics[width=1\linewidth]{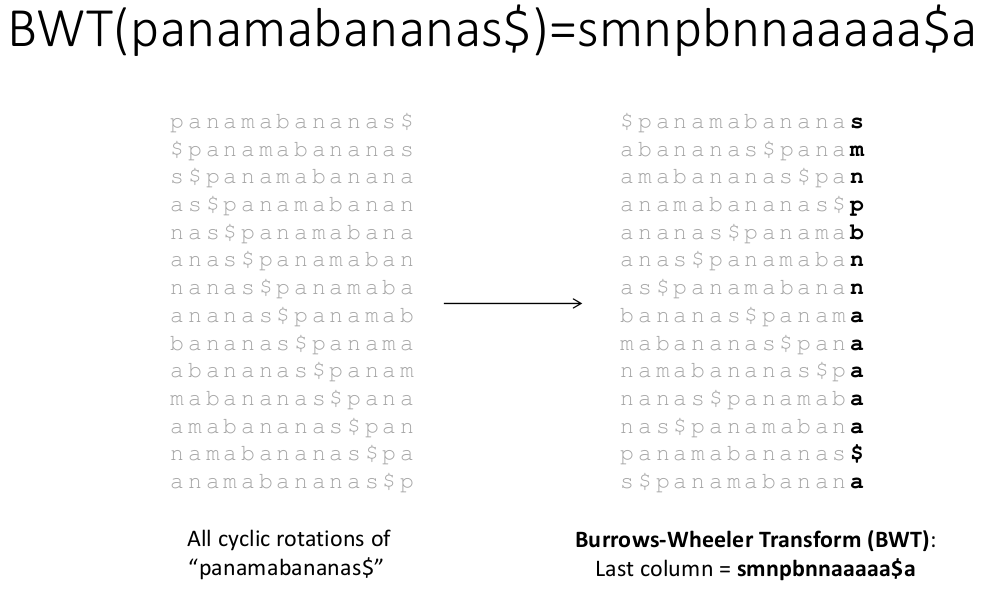}}
  \caption{Example of applying BWT on a reference sequence \cite{jones2004introduction}}
  \label{fig:BWT}
\end{figure}

Figure \ref{fig:BWT} shows an example of applying BWT on the sequence: $R=panamabananas$ \cite{jones2004introduction}. The left table shows all the rotated combinations of the reference sequence and the right table shows the table when sorted lexicographically. The last column which is bolded in the figure is the only column that needs to be stored. The last column can also be encoded using run-length encoding and stored as $s1m1n1p1b1n2a4\$a$. The compression ratio in DNA sequences is much higher than this example since there are only 4 characters which increases the probability of having long \textit{runs}. Moreover, in longer sequences, the compression reduces the required memory significantly. 

\begin{figure}[t]
  \centering
  {\includegraphics[width=1\linewidth]{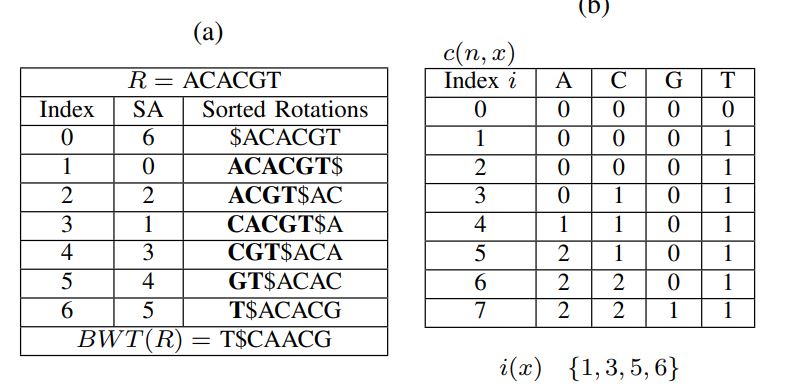}}
  \caption{Example of creating the $i(s)$, $c(n, s)$ tables from the BWT of the reference sequence \cite{ng2017reconfigurable}}
  \label{fig:BWT_Table}
\end{figure}

 To perform sequence alignment on the encoded sequence, first, the encoded sequence is loaded from memory and decoded. Then $i(s)$ and $c(n, s)$ tables are generated for the sequence. For each character \textit{s} in the reference sequence, $i(s)$ shows the index of first occurrence of the character \textit{s} in sorted BWT table \ref{fig:BWT_Table}(b). $c(n, s)$ is defined as the number of occurrences of the character \textit{s} before the index \textit{n} in the BWT of the reference, table \ref{fig:BWT_Table}(c).

To search for a query sequence in the reference sequence, first two tables are generated from the decoded sequence. Then, \textit{top} and \textit{bottom} pointer which shows the indices in which the query pattern is present are initialized to the first and last row respectively. To locate the query pattern in the reference string \textit{top} and \textit{bottom} pointers are updated recursively based on the equation \ref{eq:bwt_ipdate}.
\begin{equation}
\begin{split}
\label{eq:bwt_ipdate}
&top_{new}=c(top_{current}, s) + i(s)\\
&bottom_{new}=c(bottom_{current}, s) + i(s)
\end{split}
\end{equation}

In the following sections FPGA-based accelerators for each category are introduced.

\section{Pairwise Sequence Alignment (PSA)} \label{sec:fpga_psa} 
Although Smith-Waterman has high time complexity, $O(mn)$, it is the most common algorithm to align DNA sequences. FPGA implementations are widely developed to utilize hardware parallelism to reduce the complexity to $O(m+n)$. In one of the earliest works in accelerating SW algorithm, Lipton et al. presented a custom VLSI implementation of string alignment using a systolic array \cite{lipton1985systolic, lipton1986comparing}. The custom implementation brought orders of magnitude performance improvement to sequential computation of DNA alignment on general-purpose processors \cite{ng2017reconfigurable}. 

The works in \cite{guccione2002gene, puttegowda2003run} implement the SW algorithm on FPGA to accelerate the DNA alignment. By assuming 1 penalty for insertion and deletion and 2 for substitution, the SW equation will be simplified to equation \ref{eq:sw-simple}, where \textit{a} is the previous diagonal element and \textit{b} and \textit{c} are the non-diagonal adjacents (left and up).

\begin{equation}\label{eq:sw-simple}
d= min
\begin{cases}
a \ \ \ \ \ \ \   if (b \ or \ c = (a-1)) \ || \  R_i = S_j \\
a + 2  \ \  if (b \ and \ c = (a + 1)) \ \&\& \ R_i \neq S_j
\end{cases}
\end{equation}

\begin{figure}[t]
  \centering
  {\includegraphics[width=0.6\linewidth]{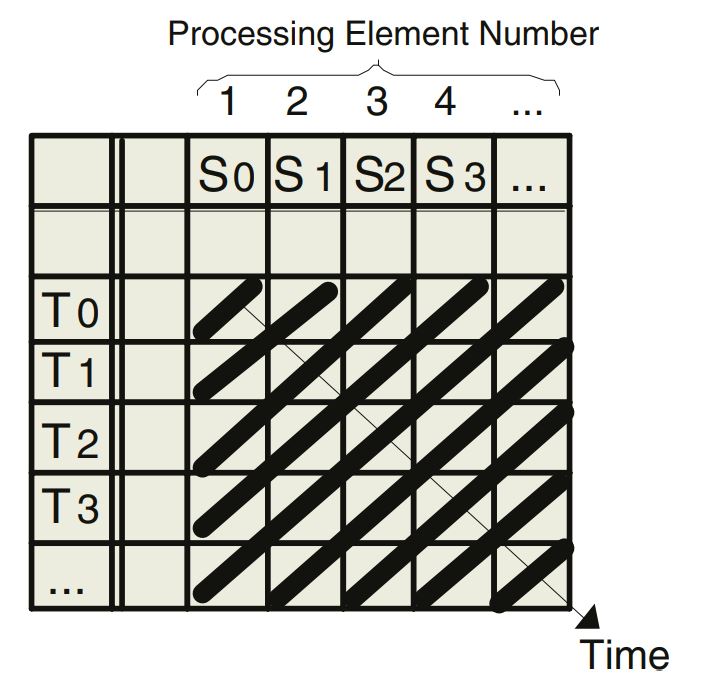}}
  \caption{Data dependencies in the alignment table, diagonal thick lanes shows the independent elements that can be computed in parallel\cite{yu2003smith}}
  \label{fig:SW_parallel}
\end{figure}

As mentioned earlier, computing each element of the alignment matrix is dependent on the computation of its up, left, and upper left elements. In figure \ref{fig:SW_parallel}, thick lines show that the diagonal elements are independent of each other and can be computed in parallel. The ability to compute the diagonal elements in parallel motivates implementing the SW algorithm on FPGAs with utilizing systolic arrays with multiple Processing Elements (PEs), each computing the score for one element of the matrix. Multiple PEs can be utilized to compute the score of independent elements. In each PE, one base of the query sequence and one base of the reference sequence are compared and the score of the aligning matrix element is computed based on the comparison. 

\begin{figure}[t]
  \centering
  {\includegraphics[width=1\linewidth]{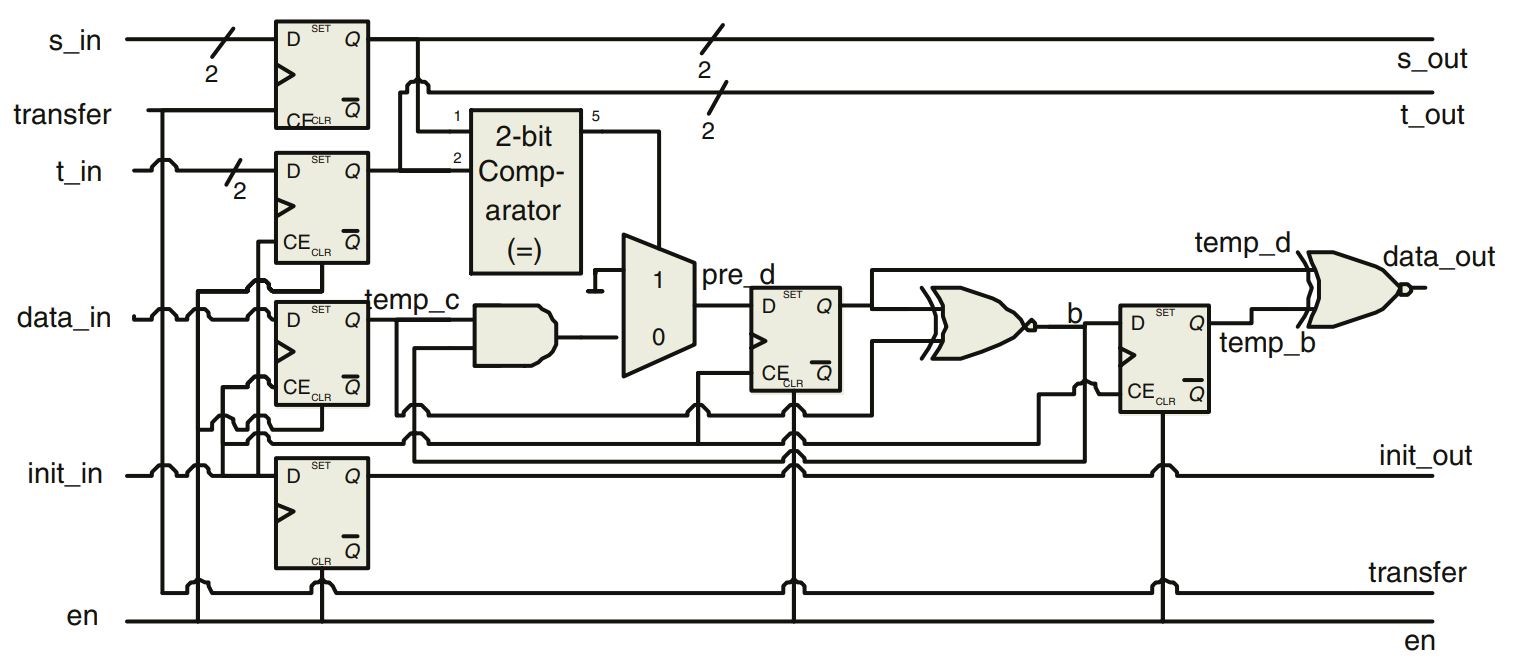}}
  \caption{Architecture of the systolic array proposed in \cite{puttegowda2003run}}
  \label{fig:PE_arch}
\end{figure}

Figure \ref{fig:PE_arch} depicts the architecture of the PEs that aligns the sequence that changes less frequently against the other one \cite{lipton1985systolic, puttegowda2003run}. The first sequence is denoted as $S$ sequence and the other is labaled as $T$  sequence. To calculate the score of element \textit{d} in equation \ref{eq:sw-simple} values of \textit{a}, \textit{b}, and \textit{c} should be available. Each PE calculate these values while comparing the characters as follows.  The value of \textit{c} is passed to the PE as data\_in input and is stored in a register. The value of \textit{b} is calculated based on the new \textit{c} value and the previous d value $b= temp\_c \ XNOR \ temp\_d$, the new \textit{b} is store in a register. The new value of \textit{d} will be the output of the multiplexer which selects between a '0' value or ($b \& temp\_c$). The output of the PE is calculated based on the new values of \textit{b} and \textit{d}, $out= temp\_b \ XNOR \ temp\_d$. This value would be the data\_in (\textit{c}) input of the next PE.

\begin{figure}[t]
  \centering
  {\includegraphics[width=1\linewidth]{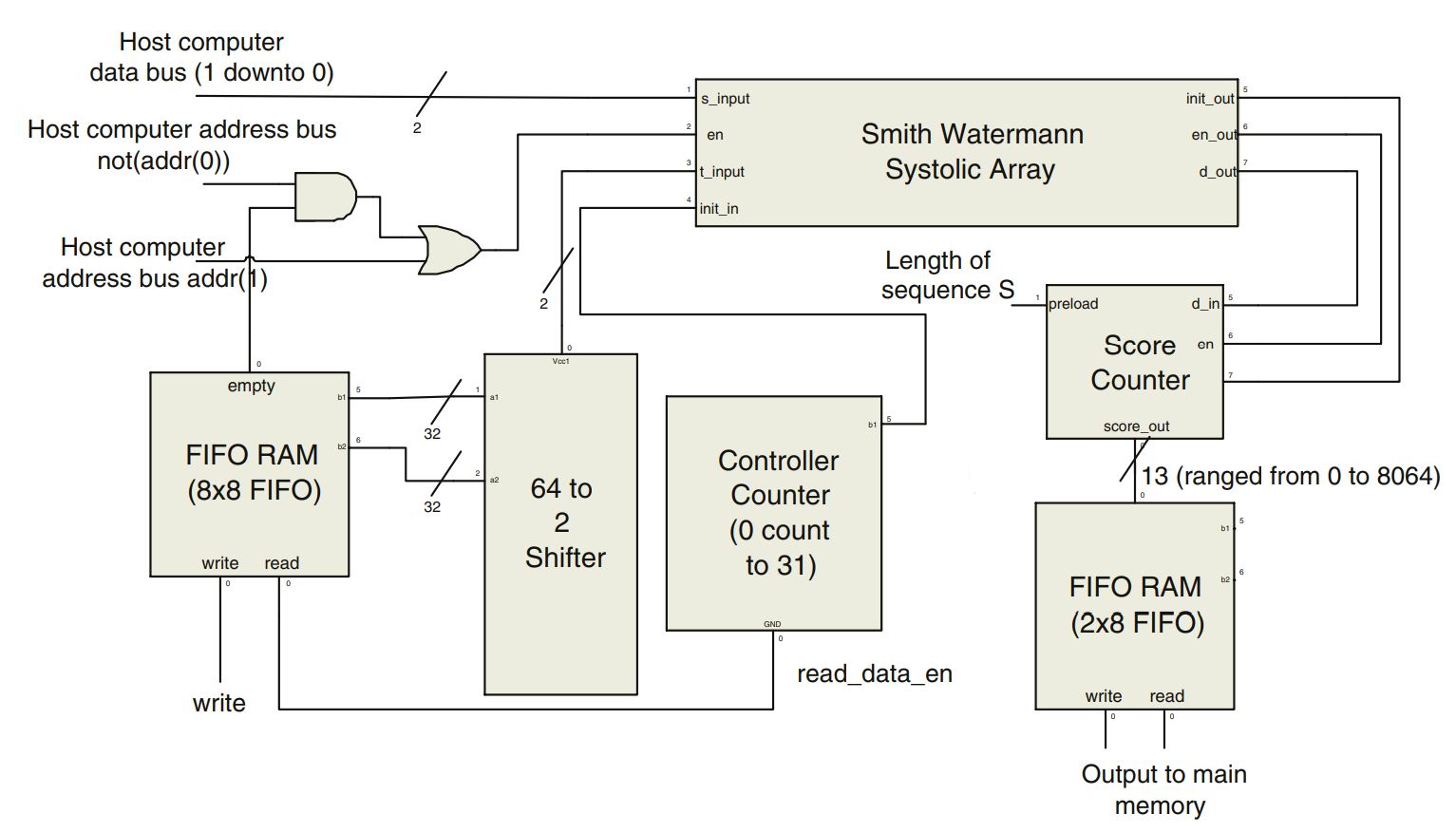}}
  \caption{Architecture of the systolic array proposed in \cite{puttegowda2003run}}
  \label{fig:sys_array}
\end{figure}

Multiple PEs are connected together to build a systolic array and the datapath of the whole system is depicted in fig \ref{fig:sys_array}. The \textit{T} sequence is streaming while the \textit{S} sequence is fixed, therefore a FIFO is used to stream the \textit{T} sequence to the systolic array. The score counter module is used to calculate the edit distance by aggregating the output of the last PE which represents the \textit{d} value in the squares of the rightmost column. When the output of the last PE is '0' then $d=b-1$ otherwise, $d=b+1$. Thus, the initialization value for the shift counter should be the length of the \textit{T} sequence and whenever the output of the last PE is 0 the counter should be decremented and otherwise it should be incremented. The final score will be achieved when the whole \textit{T} sequence is passed through the systolic array. 

In \cite{puttegowda2003run} they utilized runtime reconfiguration to fold the constant rewards (for matching) and penalties (for insertion, deletion, and substitution) into the constant adders to reduce the resource utilization of the systolic array. Additionally, instead of using general arithmetic units, one input (the query sequence) is assumed as a constant to further simplify the logic due to the constant propagation of one input. However, for a new query sequence, the FPGA systolic arrays should be reconfigured with new constant elements. The proposed systolic array architecture is depicted in figure \ref{fig:sys_array}. In this architecture, each processing element in the systolic array computes the equation \ref{eq:sw-simple} for a query element and a reference element and sends the computed output to the next processing element in the systolic array. The final output will be calculated using an up-down counter at the end of the systolic array. The counter initialization value is the query sequence length, and zero value means a perfect match.
The work in \cite{yu2003smith} implemented the systolic array with the same area and performance as \cite{puttegowda2003run} without reconfiguring the FPGA. Both query and reference sequences are loaded to the systolic array during runtime without the need for reconfiguring the FPGA. Therefore the work \cite{yu2003smith} increases the performance by eliminating the reconfiguration overhead.

Sometimes in Aligning sequences, it is desirable to have two constants penalties, one for opening a gap and one for extending the gap. Affine gap with length $l$ is defined as $gap(l)= gl+h$, where $g$ and $h$ are non-negative constants \cite{altschul1986optimal, myers1988optimal, kaya2014multiple}. The first FPGA accelerators that support the affine gap penalty were introduced in \cite{van2004families}. This paper implements a systolic array with more general processing elements. Each processing element supports different rewards and multiple penalties for gap opening and extension. It also supports both global and local alignment with supporting flexible cost function.

The work in \cite{rucci2017accelerating} exploits OpenCL and the high-level synthesis tool developed by Intel to reduce the FPGA design time while getting a better performance and power consumption than general-purpose processors. SW algorithm is implemented in OpenCL, to consider the dependencies in computing the alignment matrix, the matrix is computed in vertical blocks. Each block is computed in a row-by-row manner, starting from top to bottom, and left to right. They also evaluated the impact of using different data types in the alignment matrix on the performance and accuracy of the alignment. 8-bit \textit{char} data types shows the best performance while 64-bit \textit{long} type gives the maximum accuracy.

In \cite{rucci2018swifold} as an extension to \cite{rucci2017accelerating}, Intel Arria 10 is used as the hardware platform due to its incorporation in both new Xeon processors and the Intel-Go platform. They also provide a guide to choose the best platform for running SW for long reads. According to their results, FPGAs have been more efficient in small (less than 1M nucleotide bases, $\sim$mega-cell matrices) and medium (1M to 25M nucleotide bases, $\sim$giga-cell matrices) matrices while their efficiencies reduce as the input size gets bigger (more than 25M nucleotide bases, $\sim$tera-cell matrices). The work in \cite{zhang2007implementation} optimized the processing elements of the systolic array considering the FPGAs architecture. It uses the adder's carry output along with the combinational logic outputs in the Altera FPGAs ALM (Adaptive Logic Module) to implement the comparator used in PEs. After modification of the PE architecture, clock signals with different phases are used to create a pipeline structure with uneven stage latencies, thereby setting the clock period less than the delay of the critical path. Moreover, a compression technique is proposed to reduce the size of the block ram needed to store the alignment matrix in FPGA. All these three techniques increases the performance by fitting more PEs in FPGA and also setting the clock frequency at a higher rate.

The work in \cite{wienbrandt2013bioinformatics} used the hardware platform RIVYERA which is introduced in 2008. It has up to 128 configurable Xilinx FPGAs (Spartan3-5000). These FPGAs are distributed over 16 FPGA cards, each containing 8 FPGAs. This work focuses on short read alignment (queries with 100 base pairs), therefore each systolic array needs 100 PEs to fully parallelize the matrix calculation. Each FPGA in RIVYERA can fit up to 4 systolic arrays. In total, 512 systolic arrays are available each aligning a single query against a reference. Therefore, 512 queries can be processed concurrently. In this work, multiple FPGAs are programmed to do the same task to increase the parallelism in the query level. The other papers  such as 
\cite{houtgast2015fpga, casale2017design, wu2019fpga} 
\cite{jiang2007reconfigurable, oliver2005hyper, boukerche2007reconfigurable, gok2006efficient, faes2006scalable, li2007160}, use systolic arrays with fine-grain optimizations to support DNA alignment using SW algorithm with affine gap penalty. 

After computing the scoring matrix, to find the aligned sub-string in the reference genome, we should traceback from the highest score to a starting point. In the SW algorithm, the highest score can be anywhere in the matrix and the starting point is the first element with a score $0$. Nevertheless, in NW, due to global alignment, the highest score is among the leftmost column or the last row and the starting point is in either the first column or the first row.  Some of the accelerators rely on the host (CPU) to perform the traceback since it is not a complex and computation-intensive task. However, the work in \cite{benkrid2009highly} is parameterized to support DNA, RNA, and protein alignment, with different cost models, local/global alignment, and different lengths for the query and sequence. Moreover, this work supports traceback to output the aligned string, as well as the score matrix.

\begin{figure}[t]
  \centering
  {\includegraphics[width=1\linewidth]{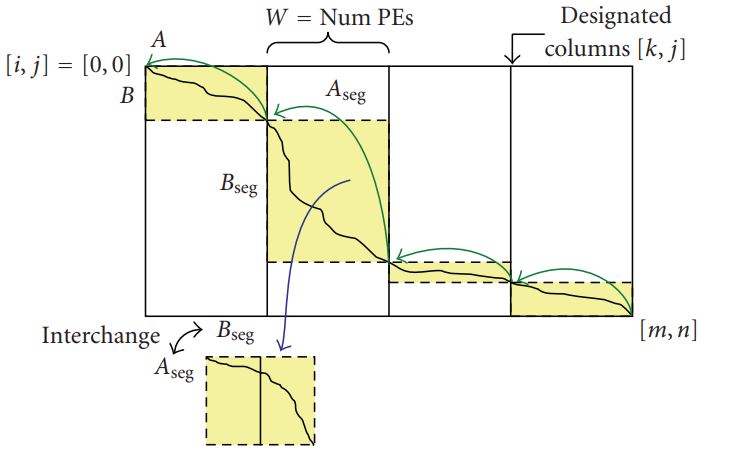}}
  \caption{Forward scan and traceback proposed in \cite{lloyd2009hardware}}
  \label{fig:SW_partioned}
\end{figure}

The work in \cite{lloyd2009hardware} proposed a global alignment accelerator with traceback support. In this work, in addition to storing the scoring matrix, each element stores the direction of the adjacent with the highest score. Also, since in current sequence alignment the length of the query sequences is higher than the number of PEs, the sequence should be portioned and processed separately. In this work each sequence is divided into sub-sequences, each has the same number of characters as the number of PEs. As depicted in figure \ref{fig:SW_partioned} each sub-sequence is aligned iteratively. To align the next sub-sequence it is enough to store the rightmost column of each sub-matrix and the final traceback can be achieved iteratively \cite{liao2018adaptively, turakhia2018darwin}.

The works in \cite{liao2018adaptively, harris2007banded} proposed a hardware acceleration of a banded SW algorithm. Since the query sequence is different from the aligned subsequent of the reference in limited elements, the solution of the alignment lays along the main diagonal strips. For instance, if only a 10\% mismatch is acceptable, the solution can be only 10\% upper or lower than the main diagonal. Figure \ref{fig:SW_banded} shows the area in the scoring matrix in which the solution should be found. Therefore, the rest of the matrix elements will not contribute to the solution and the work in \cite{liao2018adaptively} will not calculate those elements.  
\begin{figure}[t]
  \centering
  {\includegraphics[width=0.8\linewidth]{./Fig/SW_banded.pdf}}
  \caption{Aligning sequences with limited number of mismatches, deletion and insertion}
  \label{fig:SW_banded}
\end{figure}

\textbf{Comparison} \\

{\renewcommand{\arraystretch}{1.6} 
\begin{table*}[]
\resizebox{\textwidth}{!}{%
\begin{tabular}{llllllllll}
\hline
Paper                     & Year & Algorithm & Affine Gap & Flexible cost model & Traceback & Device                       & PE per device  & Frequency (MHz) & GCUPS \\ \hline
\cite{puttegowda2003run}  & 2003 & SW        & No         & No                  & No        & Xilinx Virtex II             & 7000           & 180             & 1260  \\ \hline
\cite{yu2003smith}        & 2003 & SW        & No         & No                  & No        & Xilinx Virtex XCV1000E-6     & 4032           & 202             & 814   \\ \hline
\cite{van2004families}    & 2004 & SW/NW     & Yes        & Yes                 & No        & Xilinx Virtex-II Pro XC2VP30 & 193            & 66              & 4.4   \\ \hline
\cite{lloyd2009hardware}  & 2009 & SW/NW     & Yes        & Yes                 & Yes       & Xilinx Virtex-4 FX100        & 256            & 100             & 25.6  \\ \hline
\cite{rucci2018swifold}   & 2018 & SW        & Yes        & Yes                 & No        & Intel Arria 10 GX            & N/A OpenCL HLS & N/A             & 130   \\ \hline
\cite{liao2018adaptively} & 2018 & SW         & Yes        & Yes                 & Yes       & Stratix IV                   & 128            & N/A             & 1     \\ \hline
\end{tabular}%
}
\caption{Comparing the FPGA-based accelerators of SW algorithm}
\label{tab:SW_cmp}
\end{table*}
}

During the past decade, many works have tried to accelerate implementing the SW algorithm on accelerators in general and FPGAs in particular. In the earlier works, due to sequencing issues, the queries were shorter than the current sequences. 3rd Generation of DNA sequencers produces tens of thousands of base pairs which is $100\times$ longer reads than the second generation in each run. Therefore, more flexible hardware with more general cost model is required to process these sequences.  The earliest works, e.g. \cite{puttegowda2003run, yu2003smith} were trying to fully optimize the FPGA LUTs by doing hand optimizations and their proposed hardware was limited to specific cost model without supporting the affine gap model. Although they have shown great performance, due to algorithmic limitations, they are not applicable to today's applications. Supporting a flexible cost model with the affine gap, the work in \cite{van2004families} sacrifices the performance to support a wider range of applications. Although the performance of \cite{van2004families} is two orders of magnitude less than the previous works, it is more useful due to easier hardware implementation and supporting wider variety of applications. 

The work \cite{lloyd2009hardware} implements the traceback step in FPGAs to further accelerate the SW algorithm. In the previous works, the calculated alignment matrix had to be sent to the host to traceback the scores to find the aligned substrings, however, in this work all these steps are done in FPGA to further accelerate the algorithm. With the advent of FPGA architectures and providing more resources and introducing high-level synthesis tools, the work \cite{rucci2018swifold} uses OpenCL to reduce the complexity of implementing alignment algorithm on hardware. The work \cite{liao2018adaptively} supports queries without any limitation on the length of the queries which is still $230\times$ faster than the algorithm run on the CPU. In conclusion, as time passed, more general accelerators are introduced. Although generalization reduces the performance overhead, state-of-the-arts accelerators are two orders of magnitude faster than the same algorithm run on general-purpose processors.

\section{Hash Tables} \label{sec:fpga_hash}
Heuristic algorithms (e.g. FASTA \cite{lipman1985rapid}, BLAST \cite{altschul1997gapped}, BLAT \cite{kent2002blat}, SSAHA \cite{ning2001ssaha} and etc.) are widely used in DNA alignment. All hash table based algorithms use the same seed-and-extend approach. BLAST, Basic Local Alignment Search Tools, stores the position of each Word \textit{w} (a subsequence with length $k$) of either the query or the reference in a hash table, and scans the other sequence words in the hash table. For instance all the words with length $k$ of the reference are stored in a hash table with the words as the key and the position as the content, then for each word of the query we search the hash table, if it hits, the position is an alignment candidate, called seed, otherwise for that word there is no exact match in the reference. BLAST, then extends and joins seeds, then calculates the similarity of the query with the extended seed using SW algorithm. The highest score among the extended seeds would be the output of BLAST. To support gap insertions, during the extension phase, seeds will be extended to a length greater than the query length.

The memory required to store the hash table is dependent on the length of the word and the length of the sequence. The hash table requires $2^k$ rows where $k$ is the length of the words. In each row, the occurrences of the key in the reference/query is stored. The total number of occurrences of patterns in a sequence is $|sequence|-l+k$; therefore, the total number of elements stored in $2^k$ rows of the memory is $|sequence|-k+1$. For words longer than 20 characters (40 bits) the memory required to store the reference in a hash table would be greater than the available RAM in current machines. Hence, the work \cite{homer2009bfast} proposed a two-level indexing paradigm. They build a hash table for $l$-long words where $l$ is less than $k$ (usually $l<14$). subsequences with length $l$ construct a hash table, and the rest of the $k-l$ elements are used to create the \textit{resulting bucket}. In the resulting buckets, the rest of the $k-l$ elements and their occurrence in the reference sequence is stored. To find a word with length $k$, the first $l$ elements are used as the key to the hash table and the output of the hash table is the address in the resulting bucket where the word may appear. Then a binary search is used to match the word with the contents of the resulting bucket.

Several works have tried to accelerate the BLAST algorithm on FPGAs. This algorithm consists of three main steps: I) Compiling the query into a form of a list with the length \textit{w}. II) Searching the database for exact matches (hits) between the words in the list and a substring in the reference sequence. III) In the last step, \textit{hits} are extended, and the score of aligning the extended substrings against the query is calculated. 
Technical University of Crete (TUC) architecture is proposed in \cite{sotiriades2006fpga, sotiriades2006some} that are among the earliest works tried to accelerate the BLAST algorithm on FPGAs. TUC was designed to align small queries (100 to 2000 elements) against a reference with any size using the BLAST algorithm. TUC instantiate N computing units for a query with length of N elements (2N bits). Each computing unit implements all three above steps. Each computing engine is connected to a channel to read its inputs. The first step of BLAST is precalculated before the algorithm is run, and the list is stored in the memory. Then the stream of the reference sequence is fed to the algorithm to be processed. When a hit happens, the computing unit expands the hit and calculates its score.

\begin{figure}[t]
  \centering
  {\includegraphics[width=1\linewidth]{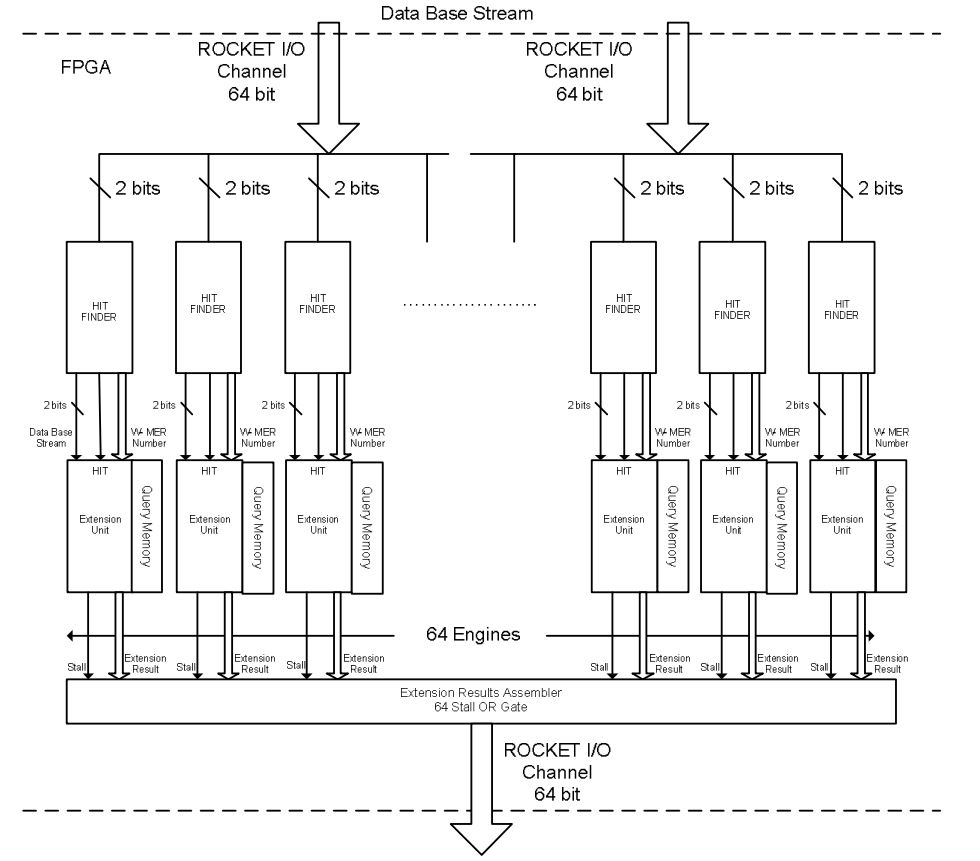}}
  \caption{TUC architecture proposed in \cite{sotiriades2006fpga}}
  \label{fig:TUC}
\end{figure}

The general architecture of TUC is illustrated in figure \ref{fig:TUC}. In database search mode, the input is the reference stream, one character for each machine. In the hash table, W-mers are used as the memory address where the positions of the W-mers in the query are stored. To reduce the memory requirement, the 12 least significant bits are used as the memory address in which the 10 most significant bits of words (W-mers) are stored in addition to the location of their appearances in the query sequence. The memory is 23bit wide, 10 bits are allocated to store the W-mer tag, 1 bit is used to show if this row is valid, and the remaining 12 bits represent the position of the W-mer in the input query. The Hit Finder Unit is responsible for implementing the second step of the algorithm. The input stream passes through an input buffer 1000 elements deep (equal to the length of queries) and is compared against the W-mer. If they are exactly matched (hit) the buffer continues to push its data to a shift register and starts to send them at the Extension Unit to compute step 3 of the algorithm. The Extension Unit, expands the hit from both sides and calculates the alignment score for each hit using SW. In the end, a hit with the highest score is the local alignment substring.

The work in \cite{sotiriades2007general} proposed a generalized platform for all five BLAST algorithms. The five different BLAST algorithms are elaborated in table \ref{tab:5blast} in BLASTn each element is nucleotide and can be represented by 2 bits. However, in BLASTp, each element is one of the 20 proteins and can be represented by 5 bits. the BLASTx, TBLASTp, and TBLASTn algorithms use translated queries or references. The translation process is representing each amino acid as three nucleotides, and each nucleotide requires 2 bits for representation; thus, each element in these three algorithms will be represented by 6 bits. The other differences between these algorithms are the scoring values in the third step as well as the length of W-mers.

The works in \cite{xia2008fpga, xia2008families, xia2008hardware}, use a systolic array to compare the W-mers with the stream of the reference sequence, instead of using a hash table to find the hits. This method reduces memory access and requirement; furthermore, it can find multiple hits in each cycle at the cost of an increase in the number of computations. In these works, the first two stages of BLAST are accelerated on FPGA while the final step is executed on the host CPU. The work \cite{chen2012reconfigurable} accelerates the first step of BLAST algorithm on FPGA. They used a bloom filter data structure to accelerate the matching stage. Bloom filter is a data structure that tells us an element either is definitely not in the set, or it may be in the set. In other words, this data structure, for a query, may output a false positive, but it never outputs a false negative. Using this data structure, that uses hashing functions to store and find elements, reduces the complexity of finding the reference substrings in the W-mers, at the cost of finding extra matches that my not truly match. These false positives will be discarded in the last step of the BLAST.

Work in \cite{bekbolat2019hblast} proposed an open-source FPGA-based accelerator that runs BLAST algorithm for queries with 256 bases and reference genomes with 2 billion base pairs.
The work in \cite{yoshimi2016accelerating} extends the acceleration to multi-FPGA platforms. They introduced a platform that consists of a PC cluster with direct connections among FPGA boards. They used a  mechanism to partition the database. Partitions are stored in the storage system which is available on the FPGA board, and they can be transmitted between FPGA boards.
{\renewcommand{\arraystretch}{1.2} 
\begin{table}[]
\centering
\caption{5 different versions of the BLAST algorithm \cite{sotiriades2007general}}
\label{tab:5blast}
\begin{tabular}{lll}
\hline
Blast Version & Query & Reference \\ \hline
Blastp & Amino acid & Amino acid \\ \hline
Blastn & Nucleotide & Nucleotide \\ \hline
Blastx & \begin{tabular}[c]{@{}l@{}}Translated \\ nucleotide sequence\end{tabular} & Amino acid \\ \hline
Tblastn & Amino acid & \begin{tabular}[c]{@{}l@{}}Translated \\ nucleotide sequence\end{tabular} \\ \hline
Tblastp & \begin{tabular}[c]{@{}l@{}}Translated \\ nucleotide sequence\end{tabular} & \begin{tabular}[c]{@{}l@{}}Translated \\ nucleotide sequence\end{tabular} \\ \hline
\end{tabular}%
\end{table}
}

The works \cite{sogabe2013acceleration, sogabe2014fpga, mcvicar2018fpga}, implemented the BFAST algorithm on FPGA. In BFAST algorithm extract seeds from the reference genome and store all of them in Candidate Alignment Locations (CALs) memory in advance with their location in the reference genome. The seeds in the query sequence (W-mers in BFAST are called seeds) are looked up in the CAL memory to find the exact matches. Then, the SW algorithm will run on the candidates to evaluate their alignment score.  In BFAST, generally the seed length $L$ is chosen 18 or 22. The seeds are divided into two parts: \textit{Addr} and \textit{Key}, with $L_a$ and $L_k$ length where $L=L_a+L_k$. In general the first 14 nucleotides in a seed are assigned to \textit{Addr}, and the rest of elements (\textit{Key}) are used in CAL table that is accessed by this \textit{Addr}. BFAST as compared to BLAST reduces the table accesses since to align each query, seeds are looked up in the CAL memory. However, this approach requires larger tables to store the reference sequence seeds and find the matched candidate efficiently. The authors of the \cite{sogabe2013acceleration} to make the algorithm more FPGA friendly, reduced the length of \textit{Addr} to 7 nucleotides. To increase the efficiency of memory accesses unlike the software approach that divides the query into multiple seeds and look up all of them immediately, they first divide the query into seeds and sort them based on their \textit{Addr}. Therefore, seeds with the same \textit{Addr} can be looked up in the CAL in parallel. In the next version of the previous work \cite{sogabe2014fpga} instead of matching seeds exactly, the proposed hardware supports one nucleotide substitution, insertion, and deletion to get higher mapping rate.

\textbf{Comparison} \\

{\renewcommand{\arraystretch}{1.6} 
\begin{table*}[]
\resizebox{\textwidth}{!}{%
\begin{tabular}{|ccccccccc|}
\hline
Paper & Year & Algorithm & Acceleration & Co-processor & Device & Query Length & Reference length & Speedup \\ \hline
\cite{sotiriades2006fpga, sotiriades2006some} & 2006 & BLAST & All stages & - & Virtex-4 4VFX140 & Long & 44M bases & 330 $\times$ \\ \hline
\cite{sotiriades2007general} & 2007 & BLAST & Stage 1, 2 & PowerPC405 300MHz & Virtex-2 Pro V2P30 & Long & - & 32 $\times$ \\ \hline
\cite{xia2008families} & 2008 & BLAST & Stage 1, 2 & Pentium 4 2.6 GHz & Stratix II EP2S130C5 & Long & 101M bases & 48 $\times$ \\ \hline
\cite{chen2012reconfigurable} & 2013 & BLAST & Stage 1 & DRC coprocessor & Virtex-5 LX330 & Long & 1.4G bases & 10 $\times$ \\ \hline
\cite{sogabe2014fpga} & 2014 & BFAST & Stage 1, 2 & Intel Core i7 2.8 GHz & Virtex-7 XC7VX690T & Short & 2.7G bases & 14 $\times$ \\ \hline
\cite{yoshimi2016accelerating} & 2016 & BLAST & Stage 1, 2 & Intel Xeon E5-2630 2.60GHz & Stratix V GX & Long & - & 3 $\times$ \\ \hline
\cite{mcvicar2018fpga} & 2018 & BFAST & All stages & Nehalem E5520 & Pico Computing M-503 & Short & 200M bases & 8 $\times$ \\ \hline
\cite{bekbolat2019hblast} & 2019 & BLAST & Stages 1, 2 & - & Virtex-7 VC709 & Short & 200M bases & N/A* \\ \hline
\end{tabular}
}
\caption{Comparing the FPGA-based accelerators of hash table-based algorithm }
\centering
\small{N/A*: The comparison result is not available in the original work}
\label{tab:BLAST_cmp}
\end{table*}
}

Heuristic algorithms such as BLAST and BFAST have been the mostly used algorithm in aligning sequences for the past decade. These algorithms are faster and more energy efficient than exact algorithms like SW/NW, while they still provide desirable coverage. As illustrated in table \ref{tab:BLAST_cmp} most of the works have tried to accelerate the first and second stages of the algorithm which are extracting seeds/W-mers and find where these seeds/W-mers are matched in the reference sequence. According to  \cite{yoshimi2016accelerating} these two steps takes 70-80\% of the execution time. Moreover, for the last step, SW algorithm calculates the score for each hit and implementing the SW on the FPGA takes a lot of FPGA resources, as mentioned in the previous section.

The early FPGA-based implementations were three orders of magnitude faster than the software implementation. This performance improvement reduced as the general purpose processors get faster with higher memory bandwidth and the recent works are less than $10 \times$ faster. Recent works \cite{sogabe2014fpga, mcvicar2018fpga} used BFAST which is an upgrade on the BLAST algorithm, this algorithm enables looking for multiple seeds at the same time and reduces the amount of computations. Integrating the host CPU with the multiple FPGA platforms, used in  \cite{yoshimi2016accelerating}, is another trend to distribute the workload among multiple execution units and achieve higher throughput and higher memory bandwidth. In general, these algorithms have reduced the amount of computation at the cost of increasing the required bandwidth as well as the number of random accesses. Any idea that reduce the amount of memory accesses and/or increase the effective bandwidth, may increase the performance and efficiency of these algorithms.

\section{BWT with FM-Index}
The human genome has $\sim$3 billion nucleotides, and storing multiple reference genomes requires a tremendous amount of storage, BWT, initially introduced as a text compression method can be used to compress the DNA sequence. BWT compression can be combined with the FM-index \cite{ferragina2000opportunistic} algorithm to align a query sequence against the compressed reference. A few hardware implementations of BWT have been proposed in the literature. Most of these platforms can solve only exact matching problems due to hardware and algorithmic limitations. The FM-index algorithm is an exact matching algorithm in nature, and supporting mismatches, insertion, and deletion in this algorithm have significant overhead on the runtime. The work \cite{fernandez2011string} proposed FHAST (FPGA Hardware Accelerated Sequencing-matching Tool), which is the first FPGA-based implementation of the BWT algorithm. BWT algorithm to align a query sequence with length $l$ requires only $l$ memory access. However, as explained in section II, the memory access pattern to update the pointers are random accesses, which has higher latency than reading consecutive memory elements. In \cite{fernandez2011string} to hide the random memory access time, multiple executing units are running concurrently and accessing the memory simultaneously. 

\begin{figure}[t]
  \centering
  {\includegraphics[width=1\linewidth]{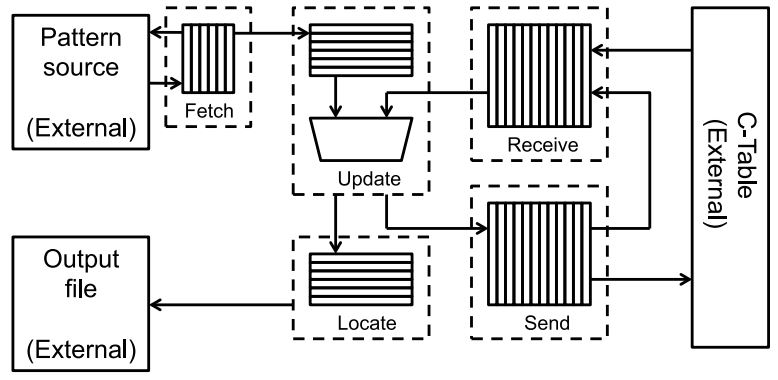}}
  \caption{Accelerating BWT algorithm on FPGA proposed in \cite{fernandez2015fhast}}
  \label{fig:BWT_FPGA}
\end{figure}

Although BWT in nature looks for exact matches, in \cite{fernandez2012multithreaded}, up to two mismatches are supported. Whenever the pattern is not matched, the unmatched character is replaced with all the three alternative characters, and the search is continued. The recent version of the work, \cite{fernandez2015fhast}, extended the platform to run multiple threads (up to 512 threads) simultaneously on a single FPGA. Each thread processes a single query sequence. 
As depicted in figure \ref{fig:BWT_FPGA}, FHAST consists of five main modules, \textit{fetch}, \textit{update}, \textit{send}, \textit{receive}, and \textit{locate}. The C-table and the list of incoming sequences are stored on the external RAM. Since the I-table is smaller it can be stored in the FPGA BRAMs. The \textit{fetch} block reads the incoming query sequences, the \textit{update} module updates the start and end pointers and finds if there is a match in the reference sequence for the query. In case of finding a match, the position of the match/matches are sent to the \textit{locate} module. The \textit{send} and \textit{receive} communicate with the off-chip memory to read the data from the C-Table.

The work \cite{arram2013reconfigurable1} takes advantage of FPGAs' dynamic reconfigurability to implement both exact and approximate sequence matching. The approximate sequence matching version of the BWT with FM-index needs much more resources as compared to the exact matching modules. In a database of short reads, approximately, $70\%-80\%$ of the short reads can be aligned using exact sequence matching. Therefore in \cite{arram2013reconfigurable1}, the FPGA is programmed with as many exact sequences matching modules as possible to align the short reads. Then the FPGA will be reconfigured with BWT approximate sequence matching modules to process the unaligned short reads in the database. The work \cite{arram2013reconfigurable2}, to reduce the number of reconfigurations, instead of using approximate sequence matching modules, it uses a pipeline of BWT with FM-index filters that support up to two mismatches. First, all the reads are processing using an exact filter, and unmatched inputs are then processed by a filter that supports one mismatch; if an input still cannot be aligned, it will be processed by the third stage that supports two mismatches. Also, they used dynamic reconfiguration to maximize the utilization of the filters by replacing the idle filters with the one that is needed.

\begin{figure}[t]
  \centering
  {\includegraphics[width=1\linewidth]{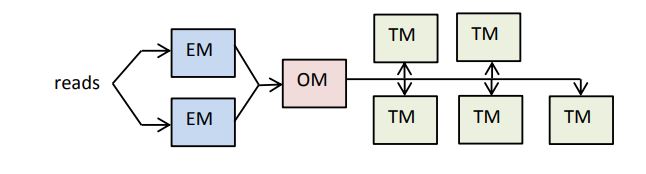}}
  \caption{Multi-FPGA design optimal configuration used in \cite{arram2015ramethy}}
  \label{fig:bwt_mfpga}
\end{figure}

Work \cite{chacon2013n} presented an algorithmic variation of the FM-index search method to reduce the number of matching iterations, called n-step FM-index. In the original BWT, the table is sorted lexicographically based on the first column, and the last column of the sorted table is stored in the memory. While, in the n-step FM-index, the table is copied for $n$ times, and the first table is sorted lexicographically based on the first column, and the $n^{th}$ table is sorted based on the $n^{th}$ column. The last columns of all the tables are stored in the memory ($n$ strings). For instance applying 2-step FM-index on the reference sequence: $acaaacatat\$$ will generate \{B[1], B[2]\} = $\{tca\$atcaaaa, aactaaa\$atc\}$ sequences\cite{chacon2013n}. In n-step FM-index, the number of iterations to match a string with length $l$ is $l/n$. The work in \cite{arram2015ramethy, arram2017leveraging} implemented the n-step FM-index on the multi-FPGA platform. Authors in \cite{arram2015ramethy} used 8 FPGAs, where they are programmed to perform each of the exact matching (EM), one mismatch (OM), or two mismatches (TM). All the incoming reads are first processed in the FPGA(s) with EM modules, and the unaligned reads are sent to the FPGA(s) with EM module. The unaligned reads in the EM FPGA(s) are sent to the final stage of FPGA(s) with TM modules. In this work to balance the execution of the stages, they used two FPGAs with EMs, one FPGA with OMs and 5 FPGAs with TMs, as depicted in fig \ref{fig:bwt_mfpga}.

\begin{figure}[t]
  \centering
  {\includegraphics[width=1\linewidth]{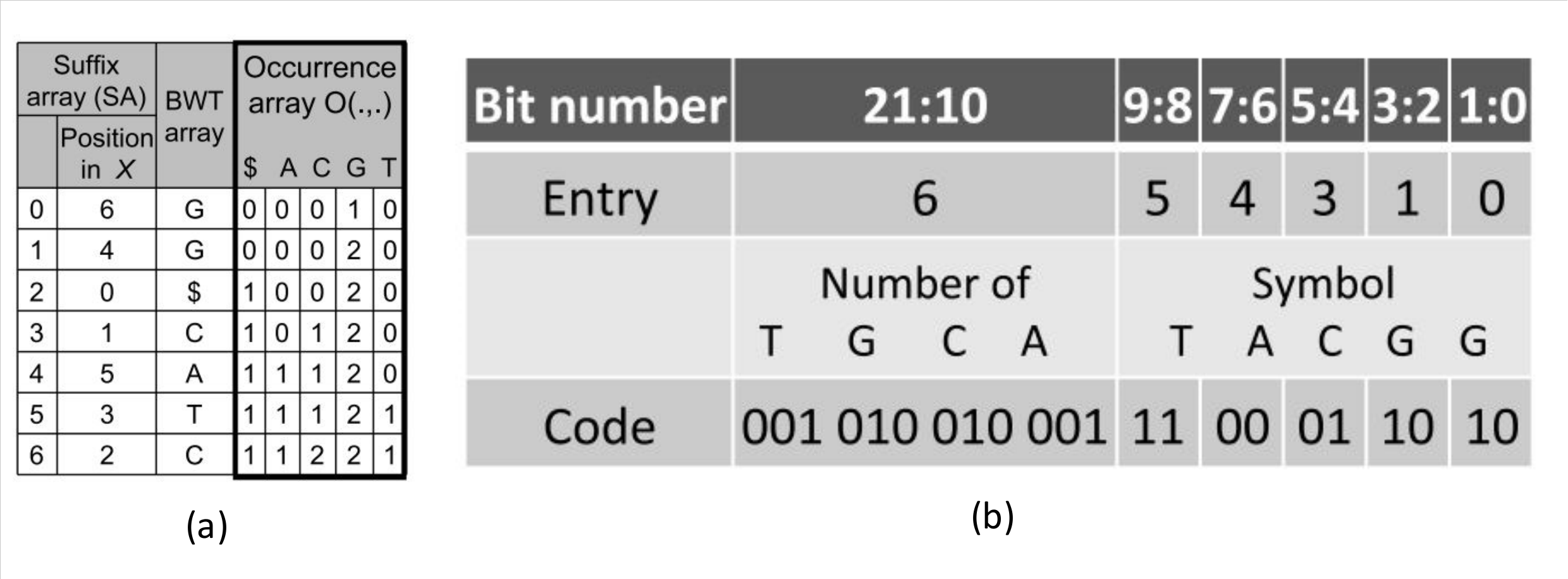}}
  \caption{Encoding method proposed in \cite{waidyasooriya2014fpga} to reduce the memory footprint of BWT}
  \label{fig:bwt_enc}
\end{figure}

The works \cite{waidyasooriya2014fpga, waidyasooriya2013implementation} proposed an encoding method that reduces the size of the required memory to store the BWT tables for 96\%.  Processing a human genome requires 48 GB of memory to store the tables, which is rarely founded in FPGA boards. To solve this issue, in \cite{waidyasooriya2014fpga}, instead of storing the occurrence table, they store the encoded table, which can be decoded in one clock cycle. Figure \ref{fig:bwt_enc}(a) shows the occurrence table needed to store the sequence $GG\$CATC$ in BWT format. While the Figure \ref{fig:bwt_enc}(b) shows the encoded format. The first $10^{th}$ bits show the first 5 elements of the sequence. The rest of 12 bits show the number of occurrences of each character.  To encode the human genome, they encode 64 elements in the occurrence array to a single encoded sequence with 256 bits. The first 128 bits show 64 elements in BWT array, while the rest of 128 bits shows the number of previously seen characters (four 32-bit numbers). In this way, only 1.5 GB of memory is required, which is 96\% less than the 48 GB required to store the occurrence table.

In BWT with FM-index algorithm, the number of iteration to align a sequence is linearly related to the length of the query; in \cite{morshedi2017fpga} to reduce the number iterations and memory accesses, the first m steps are precalculated and stored in the FPGA BRAMs. Top and Bottom pointer for all the combinations of the first $m$ characters of the query sequence ($4^m$ combination) are stored in BRAMs. When a query comes, the Top and Bottom pointers are looked up in the memory and set, then the FM-index algorithm is used to align the rest of the query elements. \\
\textbf{Comparison} \\
{\renewcommand{\arraystretch}{1.2} 
\begin{table*}[]
\resizebox{\textwidth}{!}{%
\begin{tabular}{|cccccc|}
\hline
Paper & Year & Exact/Approximate Matching & Device & Mbp/S & Speedup \\ \hline
\cite{fernandez2011string} & 2011 & Exact & Virtex-6 XC6VLX760 & 112 & 124 $\times$ \\ \hline
\cite{fernandez2012multithreaded} & 2012 & 2 mismatches & Virtex-5 LX330$\times 4$ & 59.2 & 70 $\times$ \\ \hline
\cite{arram2013reconfigurable1} & 2013 & 2 mismatches & Virtex-6 SX475T & 516 & 293 $\times$ \\ \hline
\cite{arram2013reconfigurable2} & 2013 & 2 mismatches & Virtex-6 SX475T & 97 & 18 $\times$ \\ \hline
\cite{waidyasooriya2014fpga} & 2014 & 2 mismatches & Altera EP4SGX530KH40C2 & - & 10 $\times$ \\ \hline
\cite{arram2017leveraging} & 2017 & n-step FM-index, 2 mismatches & Stratix-V $\times 8$ & 166 & 43 $\times$ \\ \hline
\cite{morshedi2017fpga} & 2017 & 2 mismatches & Virtex-6 LX240T & 78/780 & 20 $\times$ \\ \hline
\end{tabular}%
}
\caption{Comparing the FPGA-based accelerators of BWT with FM-index algorithm}
\label{tab:BLAST_cmp}
\end{table*}
}

BWT alignment is widely used in short reads due to its low latency and computational complexity. In 2011 the first FPGA-based accelerator is proposed \cite{fernandez2011string}, which could only support exact sequence matching. Although it was two orders of magnitude faster than general-purpose processors, it was not practical in real-world applications since it was limited to the exact sequence matching. Therefore, in the next version \cite{fernandez2012multithreaded} the exact matching module was followed by approximate sequence matching modules that supported up to two mismatches. Since then, all the proposed accelerators are able to support approximate sequence matching, to increase the applicability of the alignment. Supporting approximate sequence matching, exponentially, increases the computation since, for each mismatch, the number of operations multiplies by 3 (all the other nucleotides should be considered as a replacement). During time FPGA-based accelerators sacrifice the performance to become more general and more applicable. Although the work \cite{arram2013reconfigurable1} is among the fastest accelerators, it has been replaced by slower but more straightforward accelerators. This work needed dynamic reconfiguration and was not able to process unaligned sequences and postpone them to when all the reads are processed once. The work in \cite{arram2013reconfigurable2} solves this issue by using approximate sequence matching in a pipeline fashion.

The main issues in using BWT with FM-index alignment are high memory usage and irregular memory access pattern. The memory required to save the BWT tables for a human genome is 48 GB. In 2014 \cite{waidyasooriya2014fpga}, an encoding approach was proposed to reduce the memory footprint of the BWT tables. To address the latency issue, \cite{arram2017leveraging} uses n-step FM-index, which divides the latency of searching a query sequence in a reference sequence by $n$. Also, the work in \cite{morshedi2017fpga}, by pre-computing the pointers for the first $m$ elements reduces the latency along with resolving the random memory access issue. After updating the pointers for some iterations, they will get closer, and as mentioned in \cite{morshedi2017fpga}, after $\sim$9 iterations pointers will be closed enough that can be accessed in a single iteration. Thus, by using pre-computed pointers for the first $m$ elements, the memory access would become less random, which reduces the memory access overhead as well as the latency of alignment.

\section{Algorithmic Comparison}
In this section, different algorithms and their accelerators with their advantages and disadvantages have been discussed. Smit-Waterman, and Needleman-Wunsch algorithms find the optimum local and global alignments, respectively. Both algorithms support mismatches, insertion, and deletion. SW supports flexible penalties that allow the algorithm to support affine gaps. Due to its ability to find the optimum alignment, it is particularly computationally complicated with a huge memory footprint to store the alignment matrix. Several accelerators utilize the diagonal parallelization to increase the performance of alignment. However, still, after computing the alignment matrix, to find the aligned sequences, we need to traceback the matrix, which is also computationally complex.

Hash table based algorithms are significantly faster with less complexity at the cost of approximately align the sequences. However, the accuracy of the heuristic alignment is still acceptable for most of the applications. Heuristic algorithms such as BFAST and BLAST find candidates for the alignment, and they run SW algorithm in the candidates' locations to find the alignment score. Therefore, in the first step, finding candidates, these algorithms are limited to exact matching, while in the second step, calculating the alignment score using SW, they have all the advantages and disadvantages of the SW algorithms. In addition to these disadvantages, the first step requires random accesses to memory, which significantly reduces the effective bandwidth. Moreover, the memory size to store the hash tables is sometimes more than the available memory in current processing platforms.

BWT with FM-index exactly aligns the sequences, and it supports only a limited number of mismatches. However, supporting mismatches, exponentially, increases the amount of operations and, consequently, the runtime. In BWT, first, the compressed sequence should be decompressed. Then, the tables are created and stored in the memory, which is then used to update the pointers. Tables size depends on the length of the reference, which may take up to 48 GB for the human genome. However, several algorithmic updates have been proposed to reduce the memory footprint and latency. The FM-index algorithm is sequential in nature and cannot be parallelized. Nevertheless, it can be parallelized in the query level, which is aligning multiple queries against a reference sequence.
{\renewcommand{\arraystretch}{1.2} 
\begin{table*}[]
\resizebox{\textwidth}{!}{%
\begin{tabular}{|c|c|c|c|l|c|c|}
\hline
Algorithm & heuristic & Insertion/Deletion & Mismatches & Complexity & Pros & Cons \\ \hline
Smit-Waterman & No & affine gap support & Flexible penalty & O(|query|$\times|reference|$) & \begin{tabular}[c]{@{}c@{}}Supports global and local matching\\ Supports affine gap and flexible penalty\\ Diagonal elements can be parallelized\\ Regular memory accesses\end{tabular} & \begin{tabular}[c]{@{}c@{}}Computationally complex\\ Huge memory usage to store the intermediate results\\ Traceback is computationally complex\end{tabular} \\ \hline
Hash-Table & Yes & \begin{tabular}[c]{@{}c@{}}In seeds: No\\ In extension: Yes\end{tabular} & \begin{tabular}[c]{@{}c@{}}In seeds: limited\\ In extension: Yes\end{tabular} & \begin{tabular}[c]{@{}l@{}}BLAST: O(|reference|)\\ BFAST: O(|query|)\end{tabular} & \begin{tabular}[c]{@{}c@{}}Less computation complex\\ Finds candidate for SW\\ All the pros of SW in the extension phase\end{tabular} & \begin{tabular}[c]{@{}c@{}}Irregular memory accesses\\ Not flexible in seed matching\\ Huge memory usage for hash tables\end{tabular} \\ \hline
BWT with FM-index & No & No & \begin{tabular}[c]{@{}c@{}}Limited \#mismatches\\ usually 2\end{tabular} & O(|query|) & \begin{tabular}[c]{@{}c@{}}Fast exact alignment\\ Can align multiple queries against a single reference\\ Memory access pattern can be optimized\\ Stores the reference sequence in the compressed format\end{tabular} & \begin{tabular}[c]{@{}c@{}}Exact matching or limited number of mismatches\\ Exponential growth in computations w.r.t the \#mismatches\\ Random memory accesses\\ Huge memory for tables \\ Sequential in nature\end{tabular} \\ \hline
\end{tabular}%
}
\caption{Comparing three main categories of sequence alignment algorithms}
\label{tab:BLAST_cmp}
\end{table*}
}
\section{Conclusion}
During the past decade, FPGAs have been widely used in accelerating various families of DNA alignment algorithms including but not limited to Smith-Waterman, Hash-table-based algorithms, and BWT with FM-index. During the time, accelerators have become more generalized to support a wider range of applications and in some cases the performance has been sacrificed for easier and more general implementation. In this paper, we introduce the most promising works in accelerating DNA alignment algorithm families. We compare the advantages and disadvantages of different works in the same family and different families against each other.
\bibliographystyle{IEEEtran}
\bibliography{mybibliography.bib} 

\end{document}